\documentclass[twocolumn,showpacs,preprintnumbers,amsmath,amssymb,prb]{revtex4}

\usepackage{graphicx}
\usepackage{dcolumn}
\usepackage{bm} 
\usepackage{mathptmx} 
\newcounter{saveeqn}

\renewcommand{\vec}[1]{{\bf #1}}

\newcommand{\bra}[1]{\langle #1 \,|}
\newcommand{\ket}[1]{|\,#1\rangle}

\begin{document} 


\title{Quasiparticle band structure based on a generalized Kohn-Sham scheme}
\author{F. Fuchs} \email{fuchs@ifto.physik.uni-jena.de} 
\author{J. Furthm\"uller}
\author{F. Bechstedt}
\affiliation{Institut f\"ur Festk\"orpertheorie und -optik,
Friedrich-Schiller-Universit\"at, Max-Wien-Platz 1, D 07743 Jena,
Germany}
\author{M. Shishkin, G. Kresse}
\affiliation{Faculty of Physics, Universit\"at Wien, and Center for Computational Materials Science, A 1090 Wien, Austria}

\date{\today}
\begin{abstract}
We present a comparative full-potential study of generalized Kohn-Sham schemes (gKS) with explicit focus on 
their suitability as starting point for the solution of the quasiparticle equation. 
We compare $G_0W_0$ quasiparticle band structures calculated upon LDA, sX, HSE03, PBE0, and HF functionals for exchange and correlation (XC)
for Si, InN and ZnO. Furthermore, the HSE03 functional is studied and compared to the GGA for 15 non-metallic materials 
for its use as a starting point in the calculation of quasiparticle excitation energies. 
For this case, also the effects of selfconsistency in the $GW$ self-energy are analysed.
It is shown that the use of a gKS scheme as a starting point for a perturbative QP correction can improve upon
the deficiencies found for LDA or GGA staring points for compounds with shallow $d$ bands.
For these solids, the order of the valence and conduction bands is often inverted using local or semi-local 
approximations for XC, which makes perturbative $G_0W_0$ calculations unreliable.
The use of a gKS starting point allows for the calculation of fairly accurate band gaps even
in these difficult cases, and generally single-shot $G_0W_0$ calculations following
calculations using the HSE03 functional are very close to experiment.
 
\end{abstract}
\maketitle
\section{Introduction}
Density functional theory (DFT)  has become the most successful method 
for condensed matter calculations. This success is largely rooted in the 
simplicity of the exchange and correlation (XC) 
energy in the local density (LDA) or generalized gradient (GGA) approximation.
However, the underlying Kohn-Sham (KS) formalism fails in the prediction of
electronic excitation energies of semiconductors and insulators.\cite{AJW_00}
A significant step forward to correct excitation energies was achieved when the first \emph{ab initio} calculations
of quasiparticle (QP) states were performed.\cite{GSS_86,HL_85}
Their description is based on the quasiparticle equation with the XC self-energy $\Sigma$ for one excited electron 
or hole.\cite{AJW_00,GSS_86,HL_85,H_65}
In general, its solution is based on Hedin's $GW$ approximation (GWA) for the self-energy \cite{H_65} and a 
perturbative  treatment of the difference to the XC potential used in the KS equation.
The central quantity is the dynamically screened Coulomb potential $W$, which characterizes the reaction of 
the electronic system after excitation.
For many non-metals, such as the semiconductor silicon (Si), the method works well with an
accuracy of 0.1--0.3~eV for their QP gaps,\cite{AJW_00} if the Green's function $G$ is described 
by one pole at the KS (i.e., $G_0$), or better, at the QP (i.e., $G$) energy. 
However, for systems with a wrong energetic ordering of the KS bands first-order
perturbation theory is not applicable.\cite{PBODR_99}
Examples are semiconductors with a negative fundamental gap in DFT-LDA or -GGA,
e.g. InN \cite{FHFB_05} (and references therein), or with shallow $d$ bands, e.g. ZnO.\cite{RQNFS_05}

Besides the KS approach itself, the origin of the band gap problem is related to 
the semi-local approximation (LDA/GGA) for XC, which introduces an unphysical self-interaction
and lacks a derivative discontinuity.\cite{PL_83,SS_83}
These deficiencies can be partially overcome using self-interaction-free exact exchange (EXX) potentials,\cite{SMVG_97}
which are special realizations of an optimized effective potential (OEP) method.\cite{RQNFS_05}
A, by conception, different way to address the band gap problem is the use 
of a generalized Kohn-Sham (gKS) scheme, which means starting from a scheme with a spatially 
non-local XC potential.\cite{SGVML_96}
In this framework, the screened-exchange (sX) approximation uses a statically screened Coulomb kernel
instead of the bare kernel in the Hartree-Fock (HF) exchange \cite{SGVML_96}
and, with it, resembles the screened-exchange (SEX) contribution to the XC self-energy in the GWA.\cite{HL_85,H_65}
Other hybrid functionals such as those following the suggestions of Adamo and Barone (PBE0)\cite{PBE0} or 
Heyd, Scuseria, and Ernzerhof (HSE03) \cite{HSE_03} combine parts of bare or screened exchange with 
an explicit density functional. The gKS eigenvalues are usually in much better agreement with the experiment than the LDA/GGA ones. 
Therefore, the gKS solutions are supposed to be superior starting points for a QP correction,
since first-order perturbation theory should be justified. 
Hence the replacement of $G$ by $G_0$ calculated from solutions of a gKS scheme may be interpreted as a first step 
towards a self-consistent determination of the self-energy operator.\cite{KE_02,KS_02}

Here we report a systematic study of QP energies calculated from $GW$ corrections to the results of gKS schemes.
First, in section \ref{ssec:fnc} we evaluate the performance of $G_0W_0$ corrections to different gKS starting points for Si, InN, and ZnO using the
functionals sX, HSE03, PBE0, and HF. The results are compared to those of the standard KS approach based on 
an LDA functional. 
In section \ref{ssec:selfconsistent} QP gaps are calculated for a benchmark set of fifteen non-metalls utilizing the HSE03 starting point.
The effects of selfconsistency in $G$ and $W$ are discussed in  comparison to the results based on a GGA starting 
point used in Ref. \onlinecite{SK_06b}.

\section{Method}
All calculations are performed at the experimental lattice constants. 
We use the projector augmented-wave (PAW) method 
as implemented in the  Vienna {\em Ab initio} Simulation Package.\cite{KF_96}
For the details of the $GW$ implementation we refer to Refs.\ \onlinecite{SK_06} and \onlinecite{SK_06b}. 
The $GW$ calculations are carried out using a total number of 150 bands for all materials.
For the Brillouin-zone integrations, $8\times 8\times 8$  $\vec{k}$-point meshes including the $\Gamma$ point
were used, except in the case of ZnO (LDA) and InN (sX, HSE03) where the $\vec{k}$-point convergence of $W$ was found to be critical for 
meshes containing $\Gamma$. In these cases,  $8\times 8\times 8$ Monkhorst-Pack $\vec{k}$-point grids avoiding $\Gamma$ were used for
the evaluation of $W$.

One problem of the presence of shallow $d$ levels is the strong core-valence XC interaction.\cite{SDA_05,KE_02}
It can be estimated within the LDA or HF approximation, where the latter one is expected to be more reliable  
since the $GW$ self-energy approaches the bare Fock exchange operator in the short wave-length regime (i.e., at large electron binding energies). 
Therefore, we apply the HF approximation to the core-valence XC self-energy for all $GW$ calculations reported here
(see Ref. \onlinecite{SK_06b} for a validation of this approach).

Here we do not update the QP wave functions corresponding to the neglect of non-diagonal 
matrix elements of the self-energy represented in terms of the gKS wave functions $\psi^{gKS}_{\lambda}$.
The QP excitation energy $\epsilon^{\rm{N+1}}_{\lambda}$ of a state $\lambda$ in the (N+1)-th iteration is 
related the N-th iteration through the linearized equation:
\begin{equation}
\begin{split}
  \epsilon^{N+1}_{\lambda}&=\epsilon^{N}_{\lambda}+Z^{N}_{\lambda}\times\\
  &\text{Re}\left[\bra{\psi^{gKS}_{\lambda}}T+V_{n-e}+V_H+\Sigma(\epsilon^{N}_{\lambda})\ket{\psi^{gKS}_{\lambda}}-\epsilon^{N}_{\lambda}\right],
\end{split}  \label{equ_qpshift2}
\end{equation}
where $T$ is the kinetic energy operator, $V_{n-e}$ the nuclei-potential, $V_H$ the Hartree potential, and $Z^N_\lambda$ the 
renormalization factor given by
\begin{align}
  Z^N_{\lambda}=\left(1-\text{Re}\bra{\psi^{\rm{gKS}}_\lambda} 
  \frac{\delta}{\delta\epsilon}\Sigma(\epsilon)\Big|_{\epsilon^{N}_{\lambda}}\ket{\psi^{\rm{gKS}}_\lambda}\right)^{-1} .
  \label{equ_renorm}
\end{align}
The first iteration, usually denoted by $G_0W_0$,\cite{AJW_00,KE_02} is based on the gKS eigenvalues 
$\epsilon^0_{\lambda}=\epsilon^{\rm{gKS}}_{\lambda}$ as initial input to the $GW$ calculation. 
Within this approximation the perturbation operator in \eqref{equ_qpshift2} becomes 
$\Sigma(\epsilon^{\rm{gKS}}_{\lambda})-V^{\rm{gKS}}_{XC}$ corresponding to the difference between the $GW$ self-energy and the non-local 
XC potential used in the gKS equation, and the $G_0W_0$-QP shift for a certain gKS state is given by:
\begin{align}
  \Delta_{\lambda\lambda}=
  Z^0_{\lambda}\text{Re}\bra{\psi^{gKS}_{\lambda}}\Sigma(\epsilon_{\lambda}^0=\epsilon^{\rm{gKS}}_{\lambda})-
  V^{gKS}_{XC}\ket{\psi^{gKS}_{\lambda}}.
  \label{equ_qpshift}
\end{align}

\begin{table}[b]
\caption{Parameters of the gKS exchange functionals used.\label{tab_par}}
\begin{ruledtabular}
\begin{tabular}{cccc}
 Functional& $\alpha$ & $sr$ Coulomb kernel  & $\mu$ [\AA$^{-1}$]\\
\hline
    LDA    & 0.00 & $1/|\vec{x}|$                          &  -  \\
    sX     & 1.00 & $\exp(-\mu|\vec{x}|)/|\vec{x}|$            & 1.55\\
    HSE03    & 0.25 & $\rm{erfc}(\mu|\vec{x}|)/|\vec{x}|$        & 0.3 \\
    PBE0   & 0.25 & $1/|\vec{x}|$                          &  -  \\
    HF     & 1.00 & $1/|\vec{x}|$                          &  -  
\end{tabular}
\end{ruledtabular}
\end{table}
In the actual implementation of the gKS schemes, we split the gKS XC energy into the form:
\begin{align}
  E^{gKS}_{XC}=E^{DFT}_{XC}+\alpha \Big[E^{sr}_{X}(\mu) - E^{DFT,sr}_{X}(\mu) \Big], \label{equ_gKS}
\end{align}
i.e., a short-range non-local exchange term is added and treated exactly resulting
in a non-local (screened) exchange potential. The superscript $DFT$ indicates
that the respective quantity is evaluated in some (quasi)local approximation, while $E^{sr}_{X}$ 
corresponds to one of the (screened) Coulomb kernels given in Table \ref{tab_par}.
The weight $\alpha$ of the short range part and the inverse screening length $\mu$ are also listed in this table.
For simplicity, in the case of sX, the inverse screening length $\mu$ corresponding to the Thomas-Fermi wave vector 
was chosen materials independent $k_{TF}=1.55$ \AA$^{-1}$.
Gradient corrections were used for the HSE03 and PBE0 functionals.

\section{Results}
\subsection{$G_0W_0$ quasiparticle band structure of Si, ZnO, and InN}\label{ssec:fnc}
\subsubsection{Generalized Kohn-Sham bands}\label{sssec:gks_bands}
\begin{table}[t]
\caption{Direct and indirect generalized KS band gaps $E^{\rm gKS}_{g(d,i)}$
         and average $d$-band binding energies $E^{gKS}_d$ 
         calculated for cubic Si, InN, and ZnO. Experimental values are from data collections in Refs.\ 
         \onlinecite{FHFB_05,SAL_2006,Petal_05,Letal_74,Getal_97,SK_06b}. In the case of InN and ZnO they refer
         to the wurtzite polytype. All values are given in eV. \label{tab_gKS} }
\begin{ruledtabular}
\begin{tabular}{cccccccc}
     &   Energy                & LDA    &  sX   & HSE03    & PBE0 &   HF   & Exp. \\
\hline				   	          
Si   &$E^{\rm gKS}_{g,\rm i}$  & 0.51   & 0.98  & 1.04   & 1.85   & 6.57  & 1.17 \\
     &$E^{\rm gKS}_{g,\rm d}$  & 2.53   & 3.23  & 3.15   & 3.98   & 9.08  & 3.40 \\
\hline				   	          
InN  &$E^{\rm gKS}_{g}$        & -0.38  & 0.39   & 0.37  &  1.14  & 7.15  & 0.61 \\
     &$E^{\rm gKS}_{d}$        &  13.1  & 17.2   & 14.6  &  14.7  & 18.6  & 16.0-16.9 \\
\hline				   	          
ZnO  &$E^{\rm gKS}_{g}$        & 0.6    & 2.97  & 2.11   &  3.03  & 11.07 & 3.44\\
     &$E^{\rm gKS}_{d}$        &  4.6   & 8.2   & 5.7    &  5.8   & 9.3   & 7.5-8.8\\
\end{tabular}
\end{ruledtabular}
\end{table}
Results of the gKS calculations are summarized in Table \ref{tab_gKS} together with experimental results.  
We notice an increase of the computed gaps when going from the LDA to 
a truly non-local XC functional. This can be attributed to the reduction of the spurious self-interaction found in LDA, the inclusion of a potential discontinuity between filled and empty states in the XC functional,\cite{PL_83,SS_83,SGVML_96} 
and the -- in comparison to 
LDA -- enhanced core-valence exchange.  
Furthermore, for direct gaps we observe the tendency to increase from LDA over HSE03, sX, and PBE0 to the HF values.

For ZnO and InN, the $d$-band binding energies increase with respect to the LDA, approaching the experimental values.
This is an important fact to note, since the difficulties of the LDA+$G_0W_0$ approach for these compounds partly result 
from too shallow $d$ electrons in LDA, 
which is itself a result of the spuriously contained self-interaction
in local and semi-local functionals.
The too shallow $d$ electrons hybridize too strongly with the $p$ bands at the valence-band maximum (VBM),
pushing them upwards, in turn decreasing the gap ($pd$ repulsion)\cite{FHFB_05} beyond the common LDA  
gap underestimation  found for example in Si.
Contrary to the LDA, the gKS starting points yield stronger bound $d$ electrons 
(cf.\ Table \ref{tab_gKS}) with binding energies closer to the experimental values. 
Hence the gKS functionals can be expected to give a more reasonable estimate for the influence of the $pd$ repulsion on the fundamental gap
and to provide  better starting wave functions for the $GW$ calculations. 
For instance, in ZnO the Zn~$d$ character of the wave functions at the VBM is decreased 
from 0.3 for LDA to 0.25 for HSE03.

In comparison to experiment the gKS functionals, with exception of HF, generally 
perform better than standard DFT-LDA for the gaps and $d$-electron binding energies. 
For InN there is another important fact to note. All gKS functionals are found to yield the 
correct ordering of the $\Gamma_{1 c}$ and $\Gamma_{15 v}$ 
states at the zone center, in contrast to LDA findings which give a negative $sp$ gap.\cite{FHFB_05} 
This is essential for the QP description, since a correct energetic ordering of the
single-particle states is an inevitable prerequisite for a perturbative treatment of the 
$GW$ corrections.\cite{AJW_00,KS_02}

\subsubsection{Quasiparticle shifts}\label{sssec:qpshifts}

\begin{figure}[t]
  \caption{(Color online) First-order quasiparticle shifts $\Delta_{\lambda\lambda}$ versus gKS eigenvalues for
           InN (a) and ZnO (b). The valence-band maximum (VBM) and the
           conduction-band minimum (CBM) are taken as energy zeros for occupied
           and empty states, respectively. Results for five different starting
           gKS band structures are shown. 
         } \label{fig_qpshift} 
  \includegraphics[width=\columnwidth]{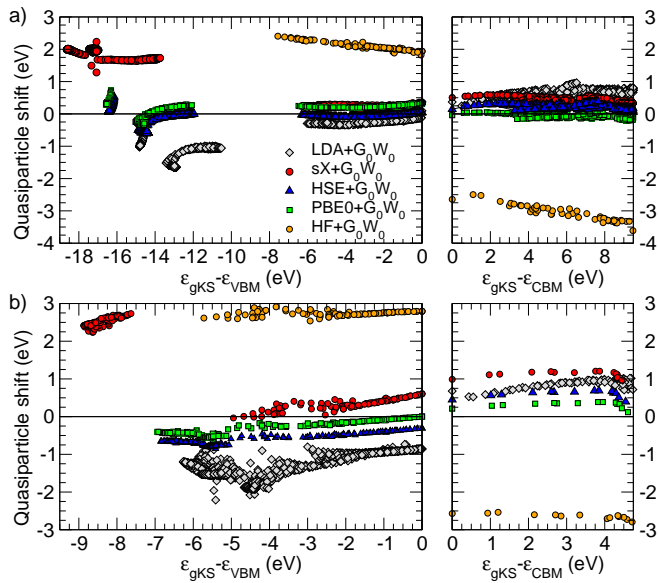}
\end{figure}

Figure \ref{fig_qpshift} shows the calculated QP shifts [Eq. \eqref{equ_qpshift}] for InN and ZnO plotted versus the gKS eigenvalues.
In the extreme limit of the HF starting point the QP shifts are given by the correlation self-energy $\Sigma_C$ 
scaled with the dynamical renormalization factor  and consequently undergo the sign change of $\Sigma_C$ at the Fermi level.
For the LDA exactly the opposite sign change is observed, {\em i.e.} valence bands aquire a negative shift
and conduction bands are shifted upwards. 
It is remarkable that the two hybrid functionals HSE03 and PBE0, which essentially mix 25~\%
HF and 75~\% DFT exchange, yield very small QP shifts. This highlights that
the 1/4 recipe is indeed a remarkable good and robust choice not only for total
energies but also for one-electron energies in semiconductors.

The performance of $GW$ upon sX is somewhat disappointing. The $d$-band positions
were very well described using the sX approach (see Table \ref{tab_gKS}), but applying
$GW$ corrections, the localized $d$ states are shifted upwards by 1.5--2~ eV,
significantly deteriorating agreement
with experiment. We will return to this issue in the next section
and in the conclusions. 
Since the SEX term in the $GW$ approximation
is very similar to the sX term in gKS schemes, the differences must be mainly related
to a different behavior of the Coulomb hole (COH) term in the $GW$ approximation and the
local-density part of the sX functional.

Turning now to a detailed discussion of the InN results [Fig.\ \ref{fig_qpshift}(a)], 
we note that the QP shifts for the upper valence bands and the conduction bands show a rather weak
dispersion with the eigenstate for all gKS schemes.
In this energy range, they may be approximated by scissors operators for occupied and empty states, whose absolute
values depend on the used gKS functional. 
Only for the lowest In~$4d$- and N~$2s$-derived bands, at gKS energies below $-10$~eV,  noticeable deviations occur
in the form of a down or upward shift for the LDA and sX starting points, respective.
In the considered energy range the QP shifts calculated for the HSE03 or PBE0 starting point 
are remarkably small with respect to the dispersion and amplitude, and they remain well below 1~eV.
For ZnO [Fig.\ \ref{fig_qpshift}(b)] similar observations can be made, however, as discussed before,
the smaller binding energy of the Zn~$3d$ states causes 
a stronger hybridization of $p$ and $d$ states than in InN. 
This might explain the upward bending of the sX shifts close to the VBM 
and the stronger bending of the LDA conduction band shifts.

\subsubsection{Non-selfconsistent QP bands} \label{sssec:G0W0}
\begin{table}[t]
\caption{Direct (d) and indirect (i) $G_0W_0$ QP band gaps, average $d$-band binding energies 
         and static electronic macroscopic dielectric constants $\varepsilon_{\infty}$ calculated upon the respective gKS 
         band structures for Si, ZnO, and InN. All energy values are given in eV. 
         Experimental results are given for comparison. \label{tab_gw} }
\begin{ruledtabular}
\begin{tabular}{cccccccc} 
     &   Energy               &  LDA &  sX  &  HSE03 & PBE0 &  HF  & Exp. \\
\hline
Si   &$E^{\rm QP}_{g,{\rm i}}$& 1.08  & 1.31  & 1.32  & 1.65   & 2.93   & 1.17 \\
     &$E^{\rm QP}_{g,{\rm d}}$& 3.18  & 3.49  & 3.48  & 3.72   & 5.21   & 3.40 \\
     &$\varepsilon_{\infty}$  & 13.9 & 10.8  & 9.8   & 7.8    & 3.4    & 11.90 \\
\hline
InN  &$E^{\rm QP}_g$          & 0.00  & 0.55   & 0.47  & 0.78   & 2.56   & 0.61  \\
     &$E^{\rm QP}_d$          & 15.1  & 15.6  & 15.2  & 15.3   & 16.6   & 16.0-16.9 \\
     &$\varepsilon_{\infty}$  & 12.2  & 6.6   & 6.8   &  4.9   & 2.4    & 7.96 \\
\hline
ZnO  &$E^{\rm QP}_g$          & 2.14  & 3.36  & 2.87  & 3.24   & 5.71   & 3.44\\ 
     &$E^{\rm QP}_d$          & 5.6   & 6.2   & 6.1   & 6.2    & 7.0    & 7.5-8.8\\
     &$\varepsilon_{\infty}$  & 5.3   & 3.0   & 3.4   & 3.0    & 1.8    & 3.74 \\
\end{tabular}
\end{ruledtabular}
\end{table}
Table \ref{tab_gw} shows the results of the $G_0W_0$ calculations for the fundamental gaps and $d$-electron binding energies.
In contrast to the LDA or gKS energies, these QP energies should have a physical meaning as 
measurable quantities and thus can be compared directly to the experimental gap values. However, it is necessary to note that
the experimental values for InN and ZnO correspond to the wurtzite instead of the zincblende polytype used in the calculations.
The zincblende gaps are expected to be about 0.2 eV smaller than the wurtzite ones.\cite{FHFB_05,SFFB_06}
For silicon, the QP gaps calculated upon the LDA, sX, or HSE03 starting point bracket the experimental values closely,  
with a maximum deviation of 0.2 eV. The LDA+$G_0W_0$ values slightly underestimate the gaps. 
In fact, this corresponds to the general trend of gap underestimation in the LDA/GGA+$G_0W_0$ approach using full-potential methods
recently established by different groups.\cite{KS_02,SK_06b}
For the gKS starting points, the calculated gaps are larger than the experimental ones. 
In more detail, starting from sX or HSE03, we find virtually identical results very close to the experiment, while 
the PBE0+$G_0W_0$ approach overestimates the gaps significantly. For the HF starting point, the largest deviations are found,
which could be expected already from  the large differences between the QP eigenvalues 
$\varepsilon^{\rm QP}_\lambda$ and the HF ones $\varepsilon^{\rm HF}_\lambda$
and the consequent break-down of the perturbative $G_0W_0$ treatment in equ. \eqref{equ_qpshift2}.
For the PBE0 starting point, however, the deviations demand for a more subtle interpretation. To understand this deviation, 
the static RPA dielectric constants calculated from the gKS eigenvalues and wave functions are included in Table \ref{tab_gw}.
Indeed, a comparison with the experimental values suggests significant underscreening for the PBE0 starting point, which certainly 
contributes to the overestimation of the gaps. 
The values calculated upon the sX and HSE03 functional are in much better agreement with the experiment, which is bracketed  closely
by them (sX/HSE03) and the LDA value.

For InN and ZnO, the trends just discussed for Si still hold, but the actual benefit of a gKS starting point becomes more apparent.
In contrast to the LDA, the gKS starting points reproduce the correct ordering of gap states for InN and yield 
more meaningful dielectric constants in the case of HSE03
and, with some restrictions, in the case of sX. Taking into account that different polytypes are compared 
the agreement between the experimental gap values and the sX+$G_0W_0$ or HSE03+$G_0W_0$ results is fair.
 
Similar to the gaps, the QP $d$-band binding energies calculated for InN and ZnO show a much weaker variation 
with the starting XC  functional than the original gKS one-electron $d$-band energies. 
The agreement of the calculated energies with experimental values (Table \ref{tab_gw}) is good for InN.  
For ZnO with more shallow $d$ levels the picture is less clear, although, here the comparison with experiment is also hampered by  
the remarkable scatter in the measured values.  
In general, we observe that the $d$ states become more strongly bound for the gKS+$G_0W_0$ approach than for  LDA+$G_0W_0$, therefore 
certainly moving in the right direction. Hence, the better treatment of exchange and correlation in the gKS starting functionals 
improves the prediction of the semi-core $d$ bands, although a general tendency towards too shallow theoretical $d$ states 
clearly remains. Note, in particular, that the $d$ bands shift upwards--- away from the experimental values  
---starting from the sX functional. Similar observations have been made for an LDA+U starting 
point in Refs. \onlinecite{Miyake06} and \onlinecite{SK_06b}. This clearly points to a deficiency
of the $GW$ approximation.

\subsection{Selfconsistent QP calculations starting from the HSE03 functional}\label{ssec:selfconsistent}
\begin{table}[t]
\caption{Results for the fundamental gaps of the HSE03 and quasiparticle ($G_0W_0$, $GW_0$ and $GW$) calculations, 
         and static electronic macroscopic dielectric constants  as used in $W_0$ (RPA). 
         The calculated values for the spin-orbit coupling (SO) induced gap-closing given in the last column have been included in the gaps.
         Also reported is the mean absolute relative error (MARE) and the mean relative error (MRE) for the gaps.
         Experimental data  for the gaps and dielectric  constants are given for comparison (for references see Ref.\ \onlinecite{SK_06b}), 
         underlined values indicate zero temperature values.
         \label{tab_bench}}
\begin{ruledtabular}
\begin{tabular}{c|c|ccc|c|cc|c}
            &  HSE03    &  $G_0W_0$&  $GW_0$   &  $GW$  &   exp.            & $\varepsilon$& $\varepsilon^{exp.}$ &SO   \\
\hline	             	
Ge          &  0.54   &   0.79   &  0.82     &  0.83   &   \underline{0.74} & 14.0  & 16.00 & 0.08 \\
Si          &  1.04   &   1.32   &  1.35     &  1.37   &   \underline{1.17} & 9.8   & 11.90 & \\ 
GaAs        &  1.12   &   1.66   &  1.71     &  1.75   &   \underline{1.52} & 9.5   & 11.10 & 0.10 \\ 
SiC         &  2.03   &   2.60   &  2.68     &  2.76   &    2.40            & 5.6   & 6.52  & \\ 
CdS         &  1.97   &   2.55   &  2.65     &  2.80   &    2.42            & 4.6   & 5.30  & 0.02  \\ 
AlP    	    &  2.09   &   2.69   &  2.77     &  2.86   &    2.45            & 6.3   & 7.54  & \\ 
GaN	    &  2.65   &   3.29   &  3.38     &  3.53   &    3.20            & 4.6   & 5.30  & 0.00 \\ 
ZnO         &  2.11   &   2.86   &  3.02     &  3.33   &   \underline{3.44} & 3.4   & 3.74  & 0.01 \\
ZnS	    &  3.05   &   3.69   &  3.79     &  3.95   &   \underline{3.91} & 4.5   & 5.13  & 0.02 \\ 
C           &  5.08   &   5.84   &  5.92     &  6.03   &   {5.48}           & 4.9   & 5.70  & \\
BN     	    &  5.54   &   6.54   &  6.66     &  6.85   &   6.1-6.4          & 3.9   & 4.50  & \\
MgO         &  6.22   &   7.94   &  8.20     &  8.66   &   7.83             & 2.6   & 3.00  & \\
LiF         & 11.2~~  &  14.1~~  & 14.5~~    & 15.2~~  &   14.20            & 1.8   & 1.90  & \\
Ar	    & 10.1~~  &  13.7~~  & 14.1~~    & 14.7~~  &   14.20            & 1.6   &  -    & \\ 
Ne	    & 14.1~~  &  20.2~~  & 20.7~~    & 21.4~~  &   21.70            & 1.2   &  -    & \\ 
\hline
MARE        & 21 \% &  6.8 \% &  8.0 \%  & 10.0 \%   \\
MRE         &-21 \% &  2.3 \% &  5.3 \%  & 9.3 \% \\
\end{tabular}
\end{ruledtabular}
\end{table}

One important question concerns the influence of selfconsistency and the resulting QP corrections in the case of a gKS starting point,
especially in comparison to local or semilocal DFT. For that reason, we have calculated the QP gaps for 15 materials, without and with
partial (only in $G$) or full (in both $G$ and $W$) selfconsistency with respect to the eigenvalues, starting from the eigenvalues and 
wave functions of the HSE03 functional. In Section \ref{sssec:qpshifts} and \ref{sssec:G0W0}, this starting point was found to give the best results for the fundamental gaps and the smallest quasiparticle shifts.
The materials adressed are non-metals, spanning the range from small-gap semiconductors to insulators. They are chosen as a 
subset of those considered in Ref.\ \onlinecite{SK_06b}. With the technical details kept largely identical, 
except for the different starting point, the data collected in Table \ref{tab_bench} 
allows for a direct and unbiased comparison of the GGA(PBE) starting point \cite{SK_06b} and the HSE03 starting point used in this work. 
The only important difference to Ref. \onlinecite{SK_06b} is that we now restore the 
all-electron charge density {\em exactly } on the plane wave grid for the calculation
of the correlation energy. This yields technically more accurate $d$-band binding energies.
Details of the applied procedure will be published elsewhere.\cite{Harl07}

Our results show that the HSE03 gaps, even though they are generally closer to the experiment than the DFT-LDA/GGA ones, 
still underestimate the experimental gaps on average by 21 \%. 
This underestimation is cured and turned into a slight overestimation of about 2.3 \% upon the inclusion of $G_0W_0$ quasiparticle corrections.
The mean absolute relative error (MARE) is reduced to 6.8 \%, which is a significant improvement compared to the 
9.9 \% MARE obtained for the GGA+$G_0W_0$ gaps.\cite{SK_06b}
Basically, this improvement results from the good performance of the HSE03 starting point for materials that comprise $d$ electrons such as
GaAs, CdS, GaN, ZnO, and ZnS, for which the HSE03+$G_0W_0$ gap-MARE is calculated to be 7.9 \%, while it is about 19.2 \% in the GGA+$G_0W_0$ approach.
We attribute the better agreement to the improved description of the $pd$ repulsion on the HSE03 level, which impacts the energy levels and 
wave functions (cf.\  Sec.\ \ref{sssec:gks_bands}). For the rest of the materials, both approaches GGA/HSE03 perform on par with a MARE of 4.8/6.1 \%.
While HSE03+$G_0W_0$ usually slightly overestimates the band gaps, the band gaps for ZnO, ZnS, LiF, Ar, and Ne remain underestimated compared
to experiment. This may be related to the large errors of the HSE03 gap exceeding 30 \% for these materials. 
Thus, the  inaccuracy of perturbation theory prevails for these systems, which share a relatively weakly screened exchange 
(static dielectric constant smaller than 4).
Another interesting point to note is the different performance of both approaches for indirect semiconductors such as Si, SiC, AlP, C, and BN.
For them the GGA+$G_0W_0$ approach performs unexpectedly well resulting in a MARE of only 2.6 \% while the performance of the 
HSE03 starting point is a little worse on average with a MARE of 8.4 \%. 

Partial selfconsistency following the $GW_0$ scheme is found to increase the gaps further by about 0.1--0.2 eV (0.4 eV for LiF, Ar, and Ne).
Since the HSE03+$G_0W_0$ gaps already showed the tendency to overestimate the experimental values, the MARE increases to 8.0 \%.
This is in contrast to the findings for the GGA+$GW_0$ approach, where an update of the eigenvalues in $G$ reduces the MARE to 5.7 \%.\cite{SK_06b}
Following the arguments given in Ref.\ \onlinecite{SK_06b}, this can be related to the electronic macroscopic dielectric 
constants $\varepsilon_{\infty}$ calculated within RPA,
which are indicative for the screening involved in $W_0$. Obviously the $\varepsilon_{\infty}$ calculated using the HSE03 functional (cf.\ Table \ref{tab_bench})
are underestimated with respect to the experiment  and, furthermore, they are generally lower than those calculated using the GGA XC-functional, 
due to the in comparison to the GGA increased gaps. 
Only inclusion of excitonic effects, i.e. electron-hole binding, allows for the calculation
of accurate electronic dielectric constants for the HSE03 functional.\cite{Shishkin07,Paier07} 

Further selfconsistent calculations, updating the eigenvalues in both $G$ and $W$, according to the $GW$ scheme, were performed.
Parallel to the findings for a GGA starting point the $GW$ gaps are larger than the $GW_0$ gaps, also for the HSE03 starting point.
Consequently they overestimate the experimental gaps with a MARE of 10 \%, which exceed the GGA+$GW$ MARE of 6.1 \%. 
The continued gap increase is found to be due to a further reduction of the screening upon updating the eigenvalues in $W$.
The difference between the GGA+$GW$ and HSE03+$GW$ schemes (updating eigenvalues in both $G$ and $W$) can 
be only related to different starting wave functions. Obviously the wave functions influence the  
resulting gap and increase it on average by 4 \%. This is a fairly small change confirming the common conjecture 
that wave functions  have only a small effect on the band gaps. Similar changes were 
observed in the self-consistent quasiparticle $GW$ (scQPGW) suggested by Faleev and Schilfgaarde.\cite{Faleev,Faleev2,Shishkin07}

\begin{table}[t]
\caption{Results for the $d$-band binding energies of GaAs, GaN, ZnO, and ZnS on different levels of quasiparticle selfconsistency compared to experimental values starting from HSE03 and PBE wavefunctions and eigenvalues, respectively.}
\label{tab_dbands}
\begin{ruledtabular}
\begin{tabular}{c|cccc|c}
       &  HSE03  & $G_0W_0$ & $GW_0$  &  $GW$  & Exp.\\
  GaAs &  17.2   &  17.5    &  17.6   &  17.6  & 18.9\\
  GaN  &  15.4   &  16.1    & 16.3    &  16.5  & 17.0\\
  ZnO  &  5.7    &   6.1    & 6.3     &  6.4   & 7.5-8.8 \\
  ZnS  &  7.5    &   7.2    & 7.2     &  7.3   & 9.0 \\
       &         &          &         &        &    \\
\hline
       &  PBE     & $G_0W_0$ & $GW_0$ &   & Exp.\\
  GaAs &  14.8   &  16.8    &  17.2   &   & 18.9\\
  GaN  &  13.3   &  15.4    & 16.1    &   & 17.0\\
  ZnO  &  5.2    &   6.1    & 6.4     &  & 7.5-8.8 \\
  ZnS  &  6.1    &   6.8    & 7.2     &  & 9.0 \\
\end{tabular}
\end{ruledtabular}
\end{table}
Finally, we address the $d$-band binding energies calculated upon the HSE03 and PBE starting point for different levels of 
quasiparticle selfconsistency,
as shown in Table \ref{tab_dbands}. We have to note that the present calculations do not include the $s$ and $p$ orbitals with the same main 
quantum number as the semi-core $d$ shell. However, since here the core-valence interaction is approximated by HF exchange 
rather than by LDA, as inherent to conventional pseudopotential calculations, the values given in Table \ref{tab_dbands} already provide a  
reasonable estimate for the $d$-band binding energies (cf.\ Ref.\ \onlinecite{SK_06} and \onlinecite{SK_06b}). 
Updated values for the PBE case are also supplied; the present values supercede those in Ref.  \onlinecite{SK_06b} and 
are more accurate, since the all-electron charge density is now accurately restored on the plane wave grid.\cite{Harl07}
In general, the QP binding energies increase over the HSE03 one-electron values due to the quasiparticle corrections, 
which can be understood mainly from the effects of the enhanced core-valence interaction in the $GW$ calculations.
Only in the case of ZnS this trend does not hold for yet unknown reasons.
However, compared to the experimental values all three quasiparticle schemes studied here underestimate the $d$-band binding energies.
With increasing selfconsistency along the row ($G_0W_0$, $GW_0$, and $GW$)  the $d$ bands shift to larger binding energies.
Thereby, for the Ga~$3d$ levels, the calculated values approach the experimental ones. 
In the case of the Zn~$3d$ levels, which are just below the $p$-like upper valence band complex, the situation is different with a more 
pronounced underestimation of the binding energies.
Analogue observations have been made for a GGA starting point,\cite{Fleszar,Miyake06,SK_06b} which 
gives a smaller $d$-band binding energy than the HSE03 starting point for each level of selfconsistency.
We note that Fleszar and Hanke observed that the inclusion of vertex corrections in the self-energy
shifts the $d$ states to stronger binding energies,\cite{Fleszar} suggesting that the neglect of such corrections 
is responsible for the erroneous behavior of the $GW$ approximation for $d$ states.

\section{Summary and Conclusions}

We have presented $G_0W_0$ QP calculations starting from a variety of XC functionals: LDA, sX, HSE03, PBE0, and HF 
for Si, InN and ZnO. We have shown that the gKS schemes, which take into account a screened exchange potential or part of it (sX, HSE03),
give  rise to eigenvalues close to the QP excitation energies. The resulting $G_0W_0$ 
corrections were found to yield QP energies in good agreement with the experimental data, and the QP shifts are less dispersive 
than those calculated upon LDA. Overall the HSE03 and PBE0 functionals gave one-electron energies very close
to the successive $GW$ calculations, resulting in small QP gap corrections across the considered 
energy range. For the HSE03+$G_0W_0$ case, the final 
QP energies were in very good agreement with experiment, whereas 
the PBE0 functional was found to yield too large QP gaps. 
We traced this back to a significant underestimation of the screening for 
the PBE0 functional, when the random phase approximation is used (also applied 
to determine $W$).

Furthermore, the QP gaps for 15 materials comprising small and large gap systems were calculated, in order to provide a 
benchmark of the HSE03 starting point against the GGA one. It was shown that the HSE03+$G_0W_0$ approach yields an 
almost halved overall error for the fundamental gaps compared to the GGA starting point. 
The largest improvement over the LDA/GGA starting points were found for materials with shallow $d$ states such as ZnO, ZnS, InN, GaAs, and GaN,
where the LDA/GGA starting point suffers from a significant underestimation of the $d$-band binding energies and a consequently overestimated repulsion between 
$p$- and $d$-like states. Since the $d$-band binding energies calculated using one of the gKS schemes (e.g. HSE03) 
are closer to the experiment, the influence of the $pd$ repulsion on the gap is described more accurately.

Furthermore, the effects of different degrees of selfconsistency were investigated. It was found that both, 
selfconsistency in $G$ ($GW_0$) and selfconsistency in $G$ and $W$ ($GW$), impair the agreement with experimental data. 
For $GW_0$, this is in contrast to the findings for a GGA starting point.\cite{SK_06b}
This could be traced back to the poorer description of the dielectric screening for the HSE03 starting point.
Selfconsistency according to the $GW$ scheme with an update of eigenvalues in both $G$ and $W$ further diminishes 
the agreement with measurements, due to a further reduction of the already underestimated screening. 
This is analoge to the findings for GGA based $GW$ calculations. 
In general, concerning selconsistent QP schemes, the present work confirms the observation already made in previous work:
to obtain accurate QP gaps it is essential to use an electronic response function and 
a screened interaction $W$ that agree closely with experiment,\cite{SK_06b,Shishkin07} 
and to combine this screened interaction  with an accurate Green's function $G$.  
For the $GW_0$ case based upon HSE03 wave functions, the overestimation of the gaps  
clearly relates to an underestimation of the static electronic screening 
employing HSE03 and the random phase approximation. 
 In this light, the success of the perturbative single-shot HSE03+$G_0W_0$ approach is a little  
bit fortuitous, since the overestimation of the screening is partially canceled 
by too small gap corrections obtained using the single-shot perturbative $G_0W_0$ approach. 
Probably the same is true for some other single shot approaches, such as 
EXX-OEP+$G_0W_0$. 
Nevertheless, if computational efficiency is an important issue--- and 
it more often is than not ---then the HSE03+$G_0W_0$ 
approach is indeed an excellent balance between accuracy  and speed.  
The calculations are as efficient as for the commonly used LDA+$G_0W_0$ method, and, with very few exceptions,  
the errors are smaller than 10~\%. 
If better accuracy is required, it can be achieved, but only by updating  
the wave functions and including excitonic effects in the calculation of the 
screening properties, i.e., vertex corrections in $W$.\cite{Shishkin07} 
For most mater, such calculations are currently not feasible due to the large
computational requirements.

A similar accuracy as for the HSE+$G_0W_0$ approach can be achieved by starting from GGA wave functions 
and eigenvalues, and updating the eigenvalues in $G$ until convergence is 
reached (see Ref. \onlinecite{SK_06b}). This approach yields comparable errors as HSE03+$G_0W_0$, but it is computationally 
more demanding, since several iterations are required to converge $G$.  
The latter method is also problematic for materials with an inverted band order in LDA/GGA. 
Which approach to choose (HSE03+$G_0W_0$ or GGA+$GW_0$) is to some extend a matter of taste, and the final results are usually 
very close and often bracket the experiment. If efficiency and robustness (band order)  
are issues, the  HSE03+$G_0W_0$ approach seems to be preferable, and it is certainly much more 
accurate than the traditional LDA+$G_0W_0$ method. 

Concerning the position of the $d$ levels, the HSE03+$G_0W_0$ method shows an underestimation of the $d$-band binding energies  
by about 1~eV  
for almost all materials. Similar observations were made for the LDA/GGA case,\cite{Fleszar,Miyake06,SK_06b}
and LDA+U based $GW$ calculations,\cite{Miyake06,SK_06b} and self-consistent quasiparticle $GW$ (scQPGW)
calculations.\cite{Shishkin07}
The underestimation of the $d$-band binding energy is thus universal to the $GW$ approximation and not related
to the starting wavefunctions.
The origin for this underestimation is yet unknown, but it is most likely related 
to the fact that the $GW$ approximation is not entirely free of self-interaction errors, and only
inclusion of vertex corrections in the self-energy might remedy this deficiency.\cite{Fleszar}

\section{Acknowledgement}
We gratefully acknowledge the support for this work by the Deutsche Forschungsgemeinschaft (Project No. Be 1346/18-2), 
the European Community in the framework of the network of excellence NANOQUANTA (Contract No. NMP4-CT-2004-500198), and  
the Austrian FWF (SFB25 IR-ON, START-Y218). Furthermore, we thank the Leibniz Rechenzentrum (LRZ) for the grants of computational time.

\end{document}